\documentclass[12pt,preprint]{aastex}

\usepackage{graphicx}

\received{}
\revised{}
\accepted{}
\cpright{AAS}{2001}
\journalid{}{}
\articleid{}{}
\ccc{}

\begin{document}

\title{A Re-interpretation of the STEREO/STE Observations and 
it's Consequences}
\author{K.~C.~Hsieh}
\affil{Department of Physics, University of Arizona,
Tucson}
\author{P.~C.~Frisch}
\affil{Department of Astronomy, University of Chicago}
\author{J.~Giacalone, J.~R.~Jokipii, J.~K\'ota}
\affil{Department of Planetary Sciences, University of Arizona}
\author{D.E. Larson, R.~P.~Lin, J.~G.~ Luhmann, Linghua Wang, }
\affil{Space Sciences Laboratory, University of California, Berkeley}

\begin {abstract}
We present an alternate interpretation of recent STEREO/STE observations 
which were attributed to energetic neutral atoms (ENA) from 
the heliosheath.   The inferred ENA intensities, as a 
function of longitude, are very similar to the instrument 
response implying that the source or sources are quite narrow.  
Such narrow sources may be quite difficult to ascribe to the available 
sources of ENA, such as the charge exchange of energetic charged 
particles with ambient neutrals, which tend to be much broader.   
We point out that the largest intensity maximum observed by STEREO/STE is centered  
at the same ecliptic longitude as the  brightest known  X-ray source, Sco X-1.  If this is  
indeed the source of the detected flux, it naturally accounts for the small 
source width.   We find that the observed energy spectrum and intensity 
are also consistent with the X-rays from Sco X-1.   If this interpretation is 
correct, then observers must take care in analyzing ENA data based on 
detectors sensitive to radiation other than ENA.    
The problem of energy dissipation in the solar wind termination 
shock remains unsolved while current understanding of the interaction 
between the solar wind and interstellar wind awaits future observations.
\end {abstract}

\keywords {helisophere, X-ray sources, energetic neutral atoms}

\section{Introduction}
Observations of energetic neutral atoms (ENA) 
from the heliosheath (the region between 
the heliospheric termination shock and the interface with the 
interstellar plasma) provide valuable constraints on the physics 
of the interaction of the Sun with the local interstellar medium.  
The recent crossings of the termination shock, first by Voyager 1 
in late 2004 (see, e.g., Decker, et al. 2005, and Stone, et al. 
2005), and more-recently by Voyager 2 in August 
of 2007 (see, e.g., Decker, et al. 2008 and Stone, et al. 2008), 
have continued a recent surge 
of activity and observations of this important part of space.  

Remote observations of photons provide additional information.   
These range from backscattered UV to radio waves.  
It has been known for more than a decade (Hsieh et al. 1992, Hsieh and
Gruntman, 1993, Hilchenbach et al. 1998) that the heliosheath 
is a significant source for ENA and that ENA can be used to 
remotely observe the heliosheath.  The recently launched IBEX is  
the first mission 
specifically intended to map the ENA emissions to study the nature 
of the interaction of the Sun with the local interstellar medium 
(McComas et al. 2004).  
Simulations and modeling of this interaction has been successful in 
accounting for many of the properties of the distant solar wind, the
termination shock and energetic particles.  

Recently, Wang, et al. (2008) published an analysis of data from the 
STE instrument on the STEREO mission.  This instrument was designed 
to measure energetic electrons from the Sun, but is also sensitive to 
neutral atoms and other radiations (Lin et al. 2008).  The instrument 
has low angular resolution in ecliptic latitude $\beta$, but its motion 
around the Sun gives it good resolution in ecliptic longitude $\lambda$.   
The hypothesis in that paper was put forth 
that the instrument was responding to ENA from the heliosheath.   

Roelof (2008) pointed out that the shapes of the flux peaks were quite 
similar to the instrumental response, and that therefore the sources 
were likely quite narrow in longitude.  Here we note that the 
location of the larger of the two 
flux maxima noted by Wang, et al. (2008, Fig. 3) is very close to 
the very bright X-ray source Sco X-1 and this is very likely the 
source of the signal.   We present arguments which support this 
hypothesis and show that the energy spectrum and intensity are also
in agreement with this interpretation.   We suggest that the secondary 
flux maximum is caused by other X-ray sources near the galactic plane.

\section {The Hypothesis of an X-ray Source}
In the following "ref. A" stands for "Wang et al. (2008)" and 
"ref. B" for "Lin et al. (2008)", 
since we will be referring to them often.

In this section, instead of interpreting the flux measure by STE-D as 
energetic neutral atoms (ENA) coming from the heliosheath, we examine 
the cause of the narrowness of the observed flux peaks reported in 
ref. A.  

When scanning a radiation field with a detector, the measured 
angular distribution of the flux is the convolution of the angular 
spread of the source and the angular response function of the instrument.  
Only for a distant point source, represented by a Dirac delta function, 
will the measured distribution reproduce the response function. 
When the full-width at half maximum (FWHM) of the two measured flux peaks 
in ecliptic longitude ($\Delta \lambda \approx 20^o$) is comparable to 
the 30$^o$ FOV of 
the individual detectors in the ecliptic, a closer look at the 
detector's angular response function is warranted. 

We examine the angular response curve of D3 of STE-D in the 
ecliptic, because D3 of STE-D on STEREO B shows the clearest 
structure of the major flux peak, afforded by the longest 
period of low solar-wind electron flux.

We examine the angular response curve of D3 of STE-D in the 
ecliptic plane, because D3 of STE-D on Stereo B shows the 
clearest structure of the major flux peak afforded by the 
longest period of low solar-wind electron flux.  Hence 
we examine its anglular response curve in the eliptic (see Figs. 
1 and 2 of ref. A).  Each detector has a 
sensitive area $A_o = (0.3)^2$ cm$^2$ and their shared rectangular 
aperture has dimensions 0.3 cm by 1.23 cm, same as the 4-detector array, 
but lying parallel to the detector plane and rotated 90$^o$ 
about the central axis connecting the center of the aperture 
and the center of the detector array.  The separation of the 
two planes is 0.889 cm.  Ignoring latitudinal effects - to be 
justified later - the angular response of D3 in the ecliptic is 
directly proportional to its exposed area projected normal to the 
incident beam from a given direction.  With this geometry, D3 has a 
30$^o$ FOV in the ecliptic spanning over $\phi = [10^o, 40^o]$, a triangular 
response function peaking at $\phi = 27^o$, and a geometrical factor of 
0.021 cm$^2$ sr.  The slightly skewed and overlapping triangular response 
curves of the four detectors of STE-U are shown in Fig. 7 of ref. B.    

The one-dimensional response function of D3 and the angular distribution 
of the major flux peak measured by D3 of STE-D on STEREO B are compared 
in Fig. 2, after normalizing the two sets to their respective maximum 
values for convenience.  The comparable angular spread in the data and 
the response function implies an extremely narrow source, which is currently very 
difficult to attribute to ENA of heliosheath origin.   The more likely 
interpretation is the detection of 3 - 15 keV X-rays from a point source 
located near $\lambda = 246^o$.  The bright and variable X-ray binary Sco X-1, 
conveniently located at ecliptic coordinates $\lambda = 245.8^o$ and 
$\beta = 5.7^o$, becomes the convincing candidate.  The 
low $\beta$ of Sco X-1 and STE-D's 80$^o$ FOV in latitude 
justify our ignoring 
any latitudinal effects in considering the response function.

To further investigate Sco X-1 as the alternative source of the major 
flux peak detected by STEREO, we plotted (crosses in Fig. 2) 
the normalized 15-50 keV X-ray flux STEREO would have detected as 
D3 scans the ecliptic longitude range 245$^o$ to 260$^o$° and 
Sco X-1 transits its FOV from 2007 DOY 159 to 188.  The X-ray flux 
used in this convolution is the daily averages for the said time 
interval, based on data from http://heasarc.gsfc.nasa.gov/docs/swift/
results/transients/.  The two data sets (dots and crosses) deviate 
from the triangular response function with similar trend, even 
thought the two sets are not congruent.  These variances can 
be understood in the context of the highly variable light curves 
of Sco X-1, a low mass X-ray binary, and in SWIFT/BAT's data coverage.  
Abrupt changes in Sco X-1's X-ray emission have durations ranging 
from minutes to weeks, but the corresponding light curves for the 
different bandpasses in the range 1.3 to 20 keV are correlated 
(McNamara et al., 1998).  Because of the frequent short-term 
variations, unless STEREO and SWIFT/BAT had identical observation times, 
the two data sets cannot track each other.  This deviation is 
worsened by the fact that the daily averages are based on varied 
data coverage, e. g., on DOY 169 (corresponds to $\lambda = 243^o$ 
in Fig. 2) SWIFT/BAT had more than 20 pointings with fluxes differing  
by a factor of seven, while on DOY 180 (corresponds to $\lambda  = 253^o$ 
in Fig. 2) there was only one pointing with a low flux. 
In view of these facts, the resemblance of the longitudinal 
distribution of the peaked "ENA" flux to that of the convolved 
X-ray flux from Sco X-1 strongly suggests Sco X-1 as the 
source of the major flux peak detected by STEREO. The variations 
among the measurements of the same flux peak by the different STE-D 
detectors on STEREO over time is consistent with this interpretation.

The extremely low-noise solid-state detectors used in STE are excellent 
X-ray detectors (Figure 5 of Tindall et al. 2008 and ref. B); but can 
Sco X-1 produce the flux and spectral shape detected by STE-D on 
STEREO A and B?  

Reanalyzing the data collected by D3 of STE-D on STEREO A in ecliptic 
longitudes $\lambda = [241^o, 252^o]$ and on STEREO B in 
$ \lambda = [238^o, 249^o] $
associated with the major flux peak assuming X-rays instead of ENA 
produces a time-averaged spectrum which can be  compared with the known X-ray 
spectra of Sco X-1.  In conversion from flux in "counts/(cm$^2$ sr s keV)" 
to a unidirectional flux in "counts/(cm$^2$ s keV)", the average 
geometrical factor of 0.025 cm$^2$ sr for four detectors 
is used.  The energy designations of the data 
points are now based on the average response of four detectors to X-rays. 

The reinterpreted STE-D measured spectrum of the major flux peak is 
compared with some published Sco X-1 X-ray spectra in Fig. 3.  The 
X-ray spectra are taken from Miyamoto et al. (1978) and Rothschild 
et al. (1980).  The model fit to a Bremsstrahlung in thermal equilibrium 
with the stellar plasma, $dj/dE = (C/E) exp(-E/kt)$, yields a temperature of 5.38 keV, which is not 
far from the range of temperature 2.5 keV (Miyamoto et al. 1978) 
to 5.15 keV (Rothschild et al. 1980), since Sco X-1 is a variable 
source.  This would explain the variations between the flux and the 
response function shown in Fig. 2 and among the fluxes registered by 
the different detectors at different times (ref. A). 

The arguments presented above, based on the angular spread of the 
major flux peak, the flux level and spectra shape, suggest strongly 
that X-rays coming from Sco X-1 is the preferred interpretation for 
the major flux peak.

\section {Discussion and Conclusions}
Noting the narrow width observed for the heliosheath ENA sources discussed 
by Wang et al. (2008), we have presented an alternate interpretation.  
We have demonstrated that the larger of the two peaks is consistent 
with it being caused by the X-ray source Sco X-1, which would produce 
a narrow peak in flux at about the correct location.  Moreover, the intensity 
and energy spectrum are also consistent with this interpretation.   Since 
producing such a narrow source using charge exchange of energetic 
charged particles and ambient neutral hydrogen is difficult, we feel 
that this identification is favored.

Judging from the shape of the minor peak at $\lambda \approx 270^o$, 
especially the negative slopes at higher $\lambda$, we believe  
this peak is also due to X-ray sources.  As Fig. 4 shows, there 
are a number of X-ray sources in the ecliptic longitude range of 
$\lambda = (260^o, 290^o$), including those in the Galactic 
Center, but none of them alone are bright enough to account for the flux 
measured in the minor peak. We tentatively identify GX5-1 at 
$\lambda = 269^o$ and $\beta = -1^o$, Sgr X4 at $\lambda = 275^o$ 
and $\beta  = -7^o$, and others 
shown in Fig. 4, as the combined source producing the minor peak.  
More detailed work will be needed to resolve the minor peak. 
We note that Collier et al. (2004) reported a similar low-energy 
ENA flux peak centered around $\lambda \approx 270^o$ (their 
Fig 4), which takes the shape of a skewed triangle with straight 
sides and a base, i.e. at zero flux, of $\approx 90$ days or 
$\Delta \lambda \approx 90^o$.  

From the experimentalist point of view, this exercise cautions 
us that X-rays are another background noise we have to deal with. 
For ENA instruments with triple coincidence, such as in HSTOF 
of CELIAS/SOHO (Hovestadt et al., 1995), HENA/IMAGE 
(Mitchell et al., 2000), and IBEX (McComas et al., 2004), 
X-ray should not be a concern.  It is very important to 
remember that all ENA images are like photon images that they 
are convolutions of the source function and the instrument 
response function.  Therefore, all observed angular distributions 
must be de-convolved prior to meaningful analysis.

This re-interpretation of the STEREO observations has consequences 
for the physics of the termination shock and heliosheath.  
The problem of the energy dissipated in the termination shock, 
suggested by Wang et al. (2008), on the basis of their original 
interpretation of the their data, remains unsolved.  Understanding of 
the dynamics and morphology of the heliosheath in the direction of 
interstellar flow remains as previously understood (e.g., Czechowski, et 
al. 2008).

\acknowledgements    
We thank R. E. Rothschild for useful suggestions and the SWIFT/BAT 
team for their help in using their data.  
This work at the University of Arizona was supported 
by NSF under grant ATM0447354 and by NASA under 
grants NNG05GE83G and NNX07AH19G, and contract 
NAS5-97271 subcontracted through 
Johns Hopkins University. The work at the University of 
California was supported under NASA grant NAS5-03131.

\clearpage

\begin{figure}
\includegraphics[width=.99\textwidth]{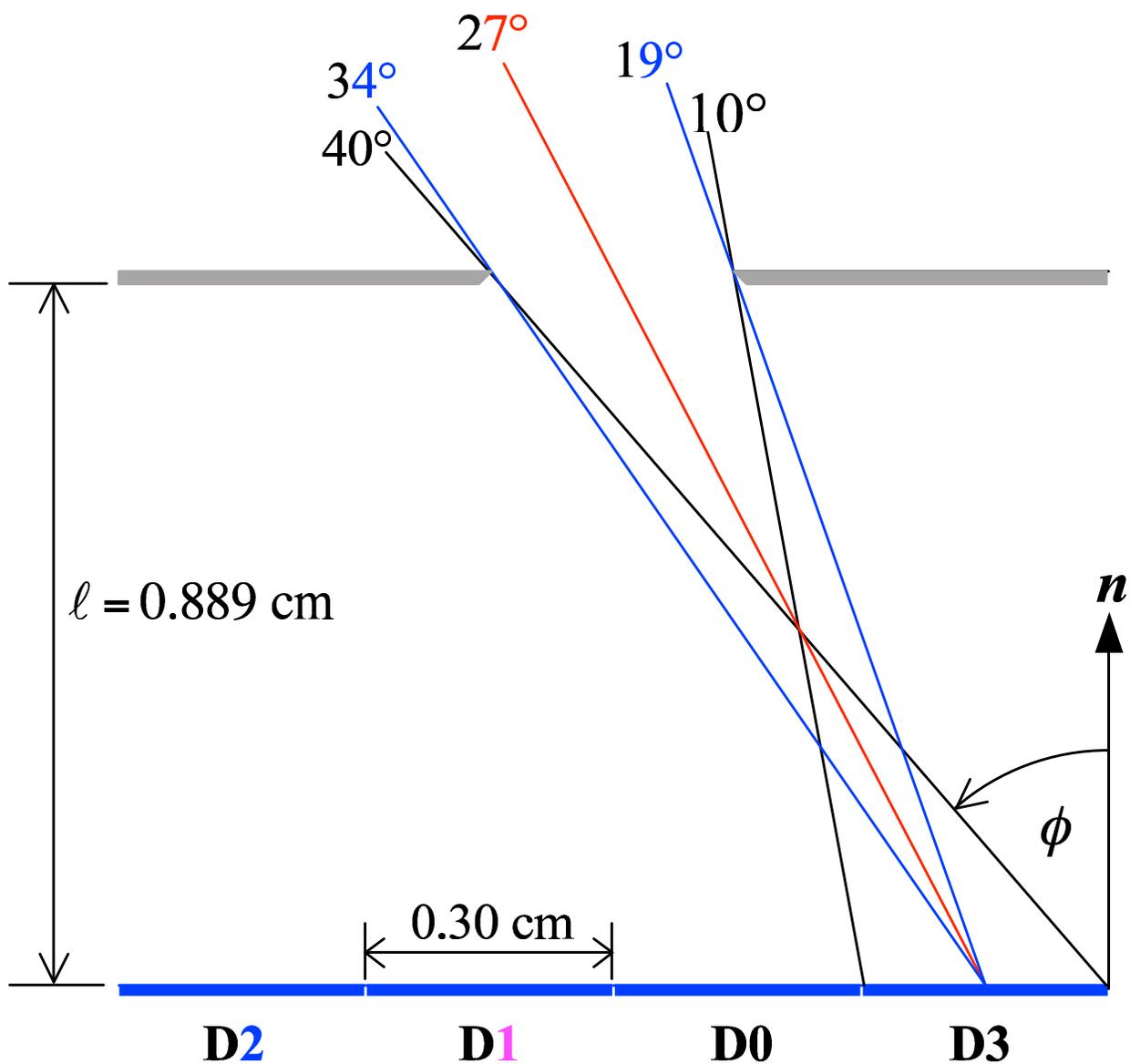}
\caption{Sketch  showing the detector-aperture geometry 
of STE-D in the ecliptic plane, also the mid-plane of the 
system (see Fig. 2 of Lin et al., 2008).  Each of the four 
detectors has area .09 cm$^2$.  The aperture has dimensions 
0.3 cm by 1.22 cm, oriented normal to the page.  Detectors 
D2 and D3 each have geometrical 
factor 0.021 cm$^2$ sr.  The entire system has a geometrical 
factor 0.10 cm$^2$ sr (Lin et al., 2008). }
\end{figure}

\begin{figure}
\includegraphics[width=0.89\textwidth]{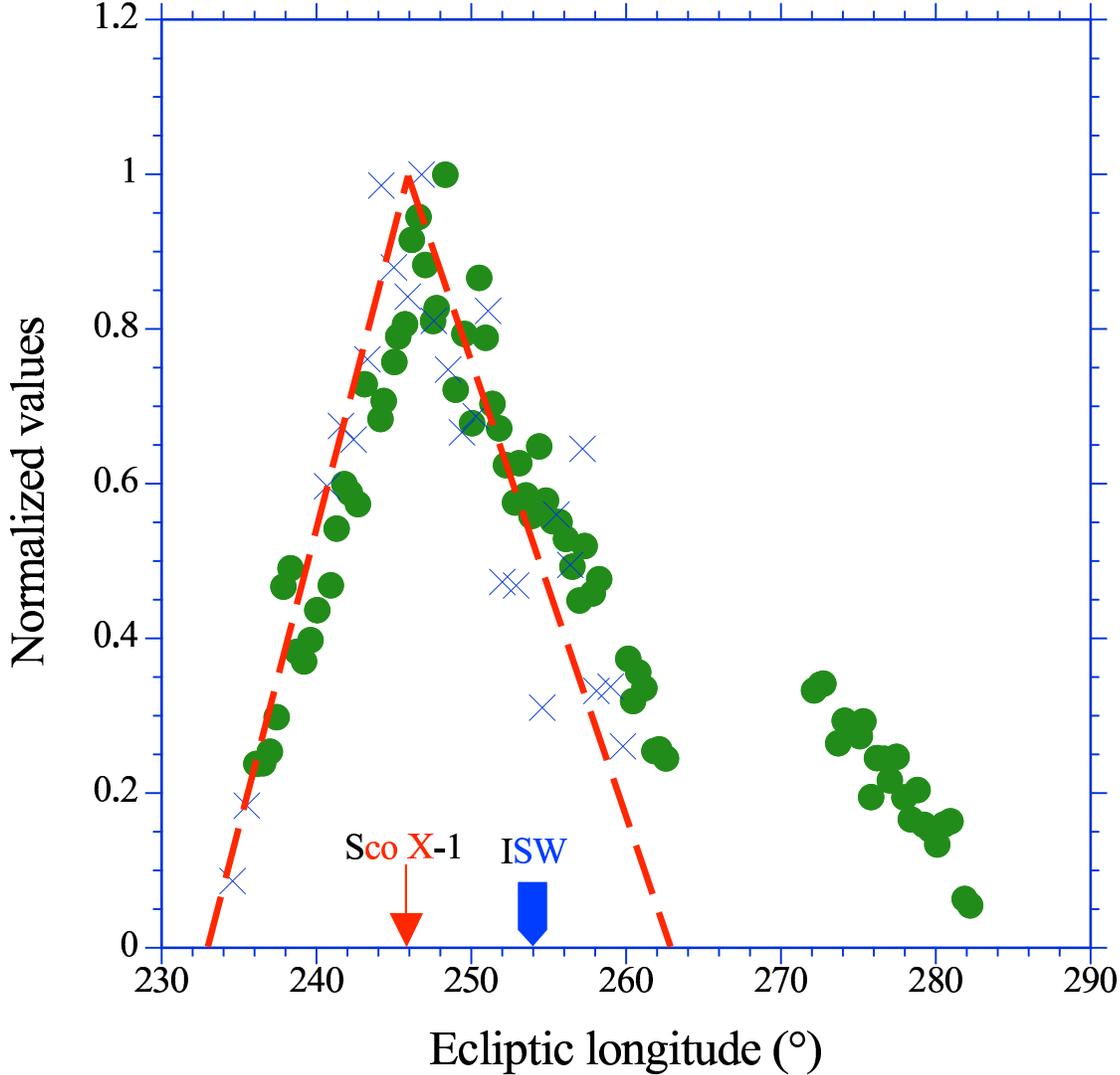}
\caption{Observed angular distributions and the angular response of 
D3 of STE-D in the ecliptic.  The dots are normalized 6.8 keV "ENA" 
flux detected by D3 on STEREO B as a function of ecliptic longitude 
(ref. A).  Overlaid is the normalized response curve of D3 
(see similar curves for STE-U in Fig. 7 of ref. B).  
The peak of this response function is set at the location of 
Sco X-1, $\lambda = 245.8^o$.  The crosses are normalized at the  
15-50 keV X-ray flux, which STEREO would have seen as Sco X-1 
transits D3's FOV. This convolution used the daily averages based 
on SWIFT/BAT data 
(http://heasarc.gsfc.nasa.gov/docs/swift/results/transients/). 
The similarity between the two data sets strongly suggests that the 
"ENA" flux may well be X-ray from Sco X-1. The deviations between the 
two sets and among the fluxes detected by the other detectors of STE-D 
are discussed in this paper.  The receding portion of the minor peak 
between $\lambda = (272^o, 283^o)$ shows a negative slope similar 
to that of the major peak, hence also suggesting X-ray sources in 
view.  The direction of the interstellar wind (ISW) is shown at 
$\lambda = 254^o$.
}
\end{figure}

\begin{figure}
\includegraphics[width=0.99\textwidth]{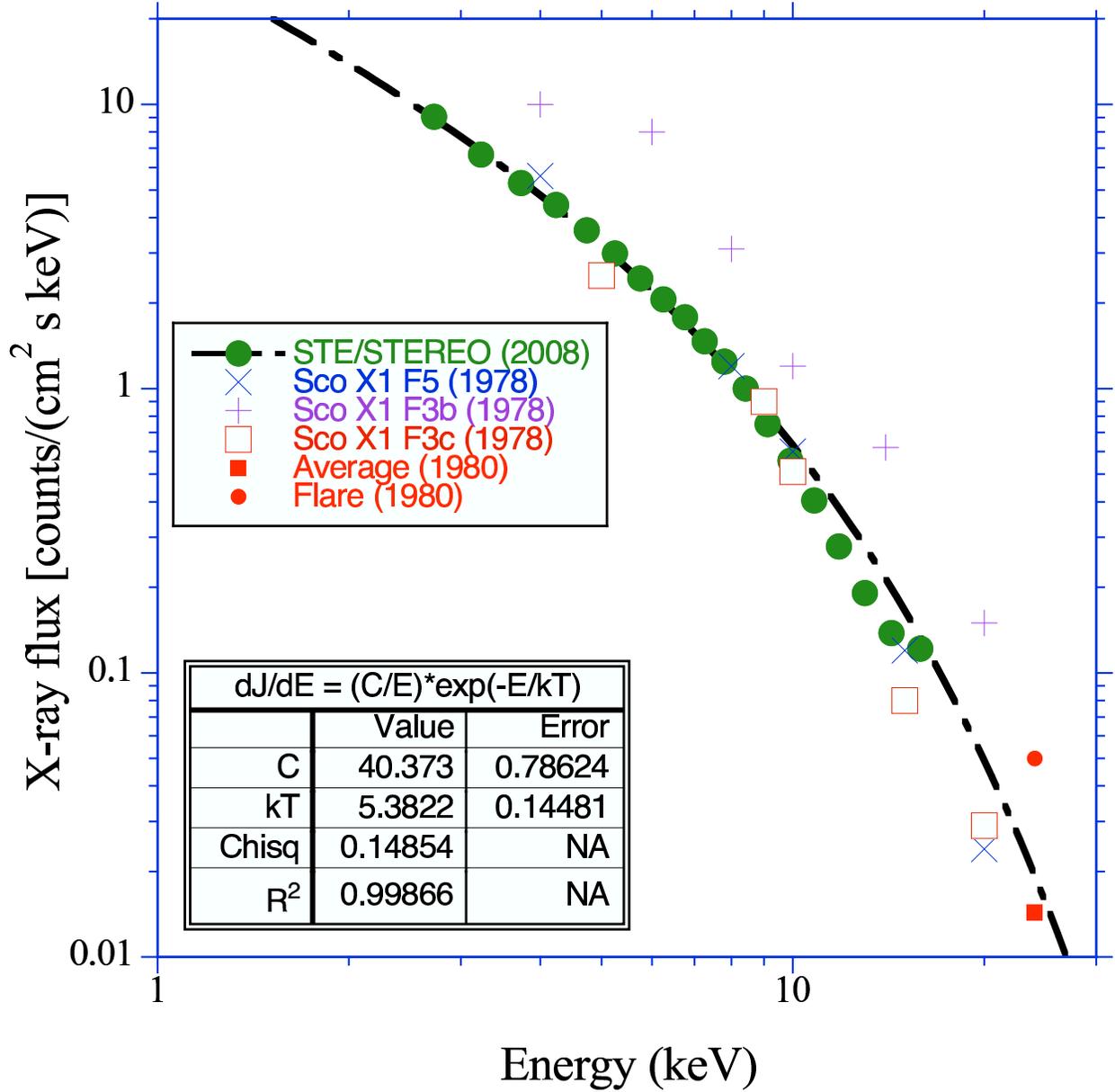}
\caption{Comparing the spectrum detected by 
STE-D/STEREO A and B 
from the major peak centered at $\lambda \approx 246^o$, 
after treating the flux as 
due to X-rays, with some published Sco X-1 X-ray spectra. The STE 
spectrum was obtained by averaging STEREO A observation over 
$\lambda = [241^o, 252^o]$ and STEREO B observations over 
$\lambda = [238^o, 249^o]$.  The
detector's X-ray calibration and geometrical factor have been used 
in constructing the spectrum as that of X-rays.  Data marked (1978) 
are taken, respectively, from Fig. 3b, Fig. 3c and Fig. 5 of 
Miyamoto et al. (1978).  The two data points at 24 keV are from 
Rothschild et al. (1980).  Sco X-1's X-ray emission is variable, 
as evident in this figure.  This would explain the variations 
shown in Fig. 2 and in ref. A.  The STE-D measurement is averaged 
over a period of time (ref. A).  The model fit to the STE-D data 
gives a temperature of 5.38 keV, not far from the range reported 
in the two references cited above.    
}
\end{figure}

\begin{figure}
\includegraphics[width=0.99\textwidth]{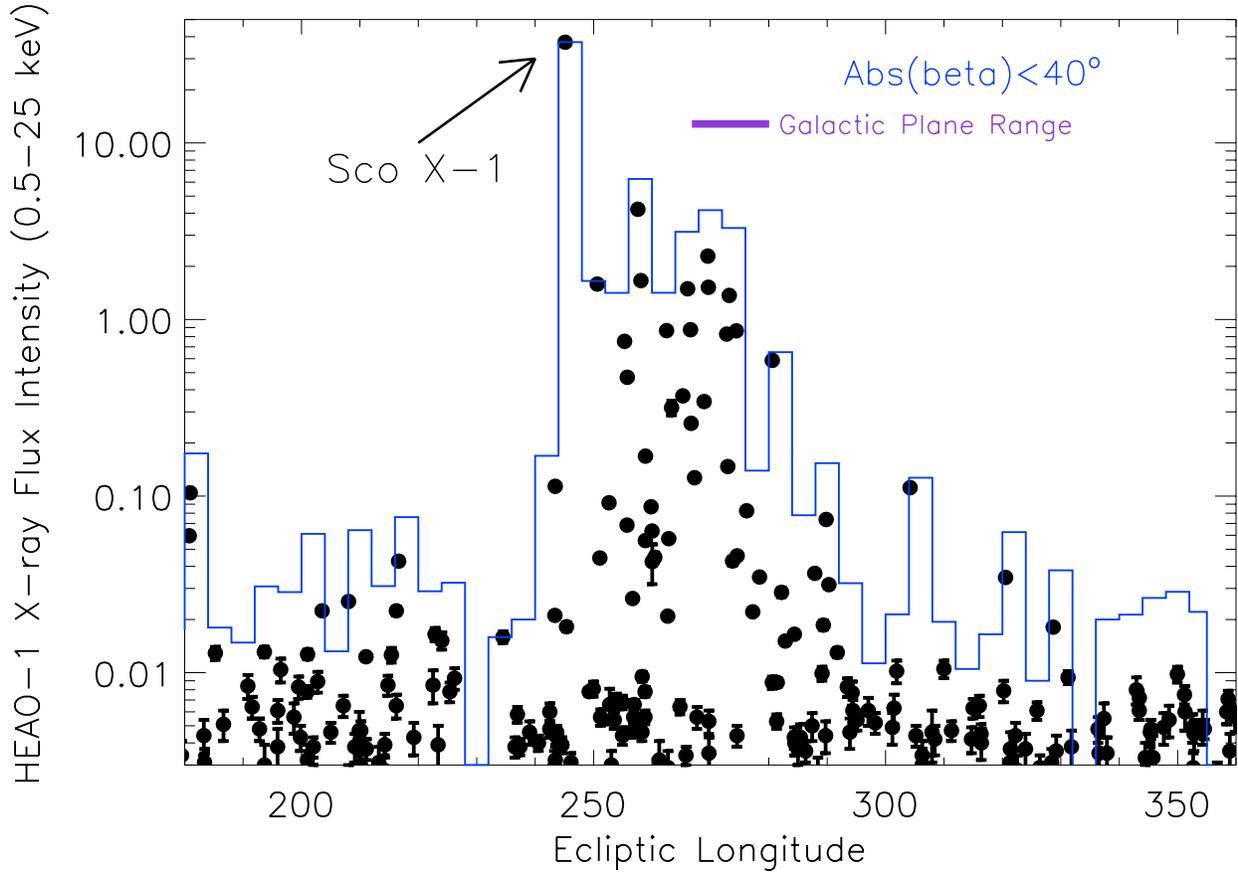}
\caption{
The brightest X-ray sources from the all-sky HEAO-1 survey 
(Wood et al. 1984).  The flux units represent the apparent 
intensity of the source in counts cm$^{-2}$ sec$^{-1}$ for photons
in the range 0.5 - 25 keV within +/- 40$^o$ of the ecliptic 
latitude $\beta$, the FoV of STE-D.  The blue histogram is the sum of
the point source fluxes, including point sources with fluxes below the
0.003 cutoff in the plot Y-axis.
}
\end{figure}


\begin{references}

\reference {} 
Collier, M. R., T. E. Moore, D. Simpson, A. Roberts, et al., 
Adv. Space Res. 34, 166 (2004).
 
  
\reference {}
Czechowski, A., M. Hilchenbach, K. C. Hsieh, S. Grzedzielski, and 
J. K\'ota, Astron. \& Astroph., 487, 329 (2008).


\reference {}
Decker, R. B., Krimigis, S. M., Roelof, E. C., Hill, M. E., 
Armstrong, T. P., Gloeckler, G., Hamilton, D. C. and Lanzerotti, L. J., 
Science, 309, 2020 (2005).


\reference {}  
Decker, R. B., S. M. Krimigis, E. C. Roelof, M. E. Hill, T. P. 
Armstrong, G. Gloeckler, and D. C. Hamilton, Nature, 
454, 67 (2008).


\reference {}
Hilchenbach, M., K. C. Hsieh, D. Hovestadt, B. Klecker, H. et al., 
 Astrophys.  J. 503, 916 (1998).


\reference{}
Hovestadt, D., M. Hilchenbach, A. Bürgi, B. Klecker, et al., Solar 
Physics, 162, 441-481 (1995).

\reference {}
Hsieh, K. C., K. L. Shih, J. R. Jokipii, and S. Grzedzielski,   
Astrophys. J.  393, 756 (1992).

\reference {}
Hsieh, K. C. and M. A. Gruntman,  Adv. Space Res.  33(6), 131-139 (1993).


\reference{}
Lin, R.\ P., D.\ W. Curtis, D.\ E.\ Larson, J.\ G. Luhmann, 
S.\ E. McBride, 
M.\ R. Maier, T. Moreau, C.\ S. Tindall, P. Turin, Linghua Wang, 
Space Sci. Rev.136, 241 (2008).

\reference {}
McNamara, R. J., T. E. Harrison, P. A. Mason, M. Templeton, 
C. W. Heikkila, T. Buckley, E. Galvan, A. Silva, and B. A. Harmon, 
Ap. J. Suppl., 116, 287 (1998).


\reference{}
McComas, D., F. Allegrini, P. Bochsler, M. Bzowski, et al., 
 Physics of the Outer 
Heliosphere: 3rd Annual IGPP Conference, eds. V. Florinski, N. V. 
Pogorelov and G. P. Zank, AIP Conference Proceedings 719, p. 
162-181 (2004).

\reference {}
Mitchell, D. G., S. Jaskulek, C. E. Schlemm, E. P. Keath, et al. , 
Space Sci. Rev. 91, p. 67, (2000).


\reference{}
Miyamoto, S., M., Matsuoka, M., M. Oda, and Y. Ogawara,  
Astron. Astrophys. 65, 329 (1978).


\reference{}
Roelof, E.\ C., Talk presented at the Huntsville Workshop 
"The Physical Processes for Energy and Plasma Transport across 
Magnetic Boundaries", Huntsville, AL, October, 2008.


\reference{}
Rothschild, R.\ E.,\ D.E.\ Gruber, F. K. Knight, P.\ L. Nolan, 
Y. Soong, Levine, A.\ M., F.\ A. Primini, W.\ A. Wheaton, 
and W.\ H.\ G. Lewin, "A high sensitivity determination of the hard X-ray 
spectrum of Sco X-1", Nature, 286, 786 (1980).

\reference {}
Stone, E. C., Cummings, A. C., McDonald, F. B., Heikkila, 
B. C., Lal, N. and Webber, W. R., 
 Science, 309, 2017 (2005).

\reference {}
Stone, E. C., A. C. Cummings, F. B. McDonald, B. C. Heikkila, N. Lal, 
and W. R. Webber,  Nature, 454, 71 (2008)
 

\reference {}
Tindall, C. S., N. P. Palaio, B. A. Ludwigt, S. E. Holland, 
D. E. Larson, D. W. Curtis, S. E. McBride, T. Moreau, R. P. Lin, 
and V. Angelopoulos,  
IEEE Trans. Nuclear Sci., 55 (2), 797 (2008).


\reference{}
Wang, L.\ , R.\  P. Lin, D.\ E. Larson, and J.\ G. Luhmann ,  
Nature, 454, 81 (2008).

\reference {}
Wood K.S., Meekins J.F., Yentis D.J., Smathers H.W., McNutt D.P.,
    Bleach R.D., Byram E.T., Chubb T.A., Friedman H., Meidav M.,
   Astroph. Jour. Suppl. 56, 507 (1984).
   



\end{references}
\end{document}